\documentclass[]{spie}

\def\msun{M$_{\odot}$}

\usepackage[english]{babel}
\usepackage[utf8x]{inputenc}
\usepackage{amsmath}
\usepackage{graphicx}
\usepackage[colorinlistoftodos]{todonotes}
\usepackage{hyperref}

\title{Grown-up stars physics with MATISSE}

\author{F. Millour\supit{a,}\footnote{\ \ O. Chesneau coordinated the
    stellar physics science group of MATISSE until his demise in June
    2014.},
  J. Hron\supit{b},
  A. Chiavassa\supit{a},
  G. Weigelt\supit{c},
  A. Soulain\supit{a},
  Z. Khorrami\supit{a},
  A. Meilland\supit{a},
  N. Nardetto\supit{a},
  C. Paladini\supit{d},
  A. Domiciano de Souza\supit{a},
  G. Niccolini\supit{a},
  K.-H. Hofmann\supit{a},
  D.~Schertl\supit{a},
  P. Stee\supit{a},
  P. Bendjoya\supit{a},
  F. Th\'evenin\supit{a},
  F. Vakili\supit{a},
  P. Berio\supit{a},
  T. Lanz\supit{a},
  A. Matter\supit{a},
  P. Cruzal\`ebes\supit{a},
  R. Petrov\supit{a}, 
  B. Lopez\supit{a}.
  \skiplinehalf
  \supit{a}Universit\'e C\^ote d'Azur, OCA, CNRS, Lagrange, France; \\
  \supit{b}University of Vienna, Department of Astrophysics, Vienna, Austria;\\
  \supit{c}Max Planck Institute for Radio Astronomy, Bonn, Germany;\\
  \supit{d}Institut d'Astronomie et d'Astrophysique, Universit\'e Libre de Bruxelles, Brussels, Belgium.
}

\authorinfo{
\vspace{0.cm} \\\emph{Further author information:}\\ Send correspondence to F. Millour, Lagrange, OCA, Bd de l'Observatoire, CS 34229, 06304, Nice, France\\E-mail: \href{mailto:fmillour@oca.eu}{fmillour@oca.eu}, Telephone: +33 (0)4 92 00 30 68, or +33 (0)4 92 07 64 89}

\begin{document}
\maketitle


\begin{abstract}
MATISSE represents a great opportunity to image the environment around massive and evolved stars. This will allow one to put constraints on the circumstellar structure, on the mass ejection of dust and its reorganization, and on the dust-nature and formation processes. MATISSE measurements will often be pivotal for the understanding of large multiwavelength datasets on the same targets collected through many high-angular resolution facilities at ESO like sub-millimeter interferometry (ALMA), near-infrared adaptive optics (NACO, SPHERE), interferometry (PIONIER, GRAVITY), spectroscopy (CRIRES), and mid-infrared imaging (VISIR). Among main sequence and evolved stars, several cases of interest have been identified that we describe in this paper.
\end{abstract}

\section{How many evolved stars can be observed with MATISSE?}

First of all, we need to emphasize that MATISSE will mainly focus on dusty stars, as the mid-infrared is the perfect match to collect data on the dust quantity and composition around a given object.

To illustrate that, we selected a few topics of interest that will be developed in this paper, and counted the number of stars that will be observable with VLTI/MATISSE. We included a quantity of non-dusty targets (regular WR stars and Be stars) in the sample to compare the performances of MATISSE with its prime targets. To do so, one need to take into account the sensitivity of the instrument itself in the L- and N-bands\cite{Matter2016}, of course, but also the VLTI infrastructure subsystems sensitivity, which can be sometimes the limiting factor, especially for red targets. We need therefore to take into account: the telescope guiding limit in the V-band (V=13.5 and V=17 for ATs STRAP tip-tilt and UTs MACAO AO, respectively), and the fringe tracker limit in the K-band for GRA4MAT and other devices like IRIS (K=7.5 for ATs and K=10 for UTs).

We considered star lists on Wolf-Rayet stars (WR), Red supergiant stars (RSG), Asymptotic Giant Branch stars (AGB), Be stars and B[e] supergiant stars visible from Paranal ; we extracted them from the following catalogs:
\begin{itemize}
\item Rosslowe \& Crowther (2014) for WR stars\cite{2015MNRAS.447.2322R},
\item Hoffleit (1991) for RSG stars\cite{1991bsc..book.....H}, 
\item we scanned through ADS publications on the topic for AGB stars,
\item Frémat et al. (2005)\cite{2005A&A...440..305F} and Yudin (2001)\cite{2001A&A...368..912Y} for Be stars, 
\item For B[e] supergiant stars we used our own catalog of stars.
\end{itemize}

When available, we retrieved the SED of each object  (V, K, L and N magnitudes) and ISO/IRAS spectra (if it exist).
Magnitudes of these targets were compared with the theoretical limits of MATISSE, considering spectrally-usable data, i.e. data with a SNR$\geq3$ per spectral channel during the expected optimal DIT and exposure time combination, with and without fringe tracker\cite{Matter2016}.

\begin{figure}[htbp]
  \begin{center}
    \begin{tabular}{ccc}
      \includegraphics[width=.47\textwidth]{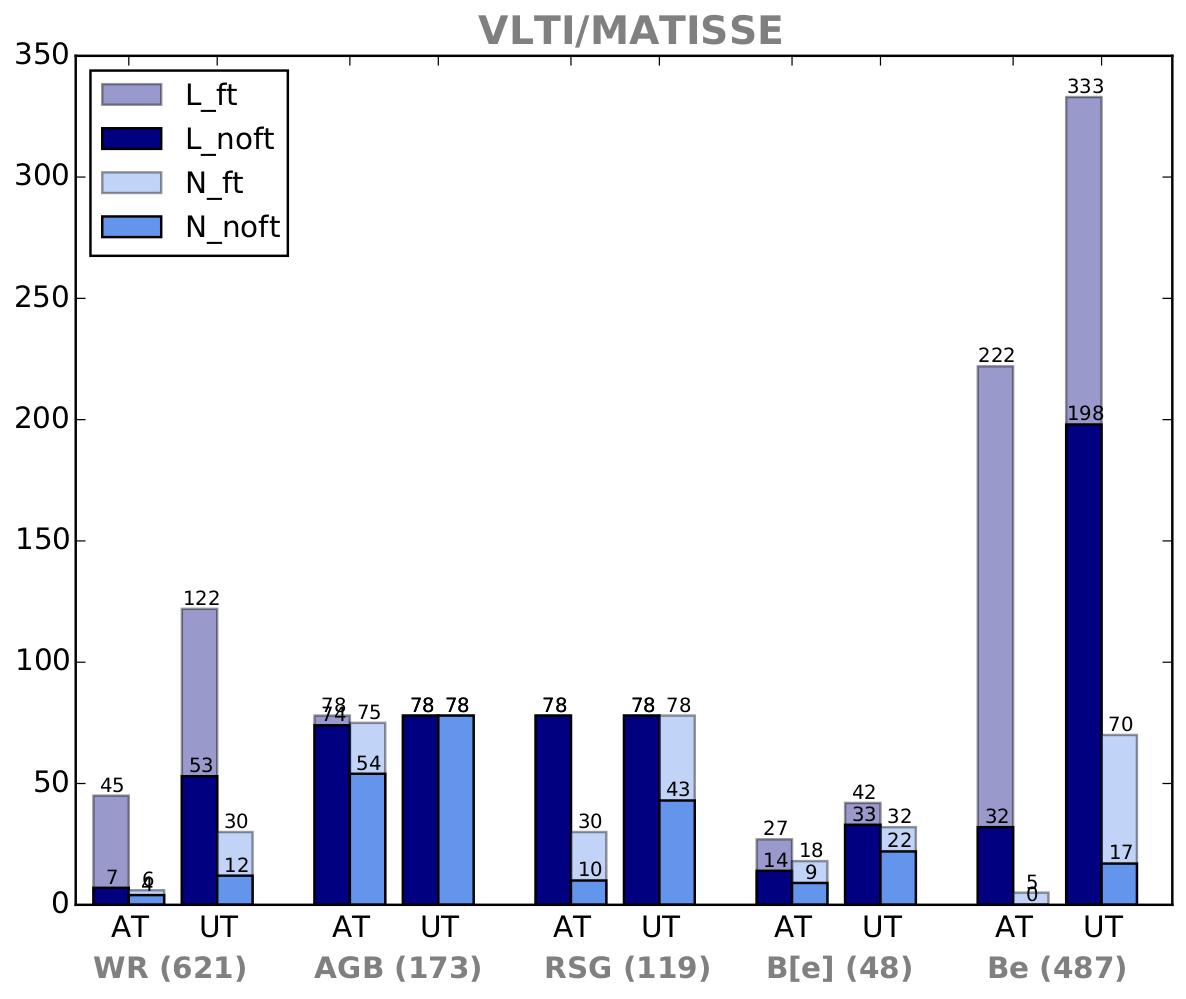}&
      \includegraphics[width=.47\textwidth]{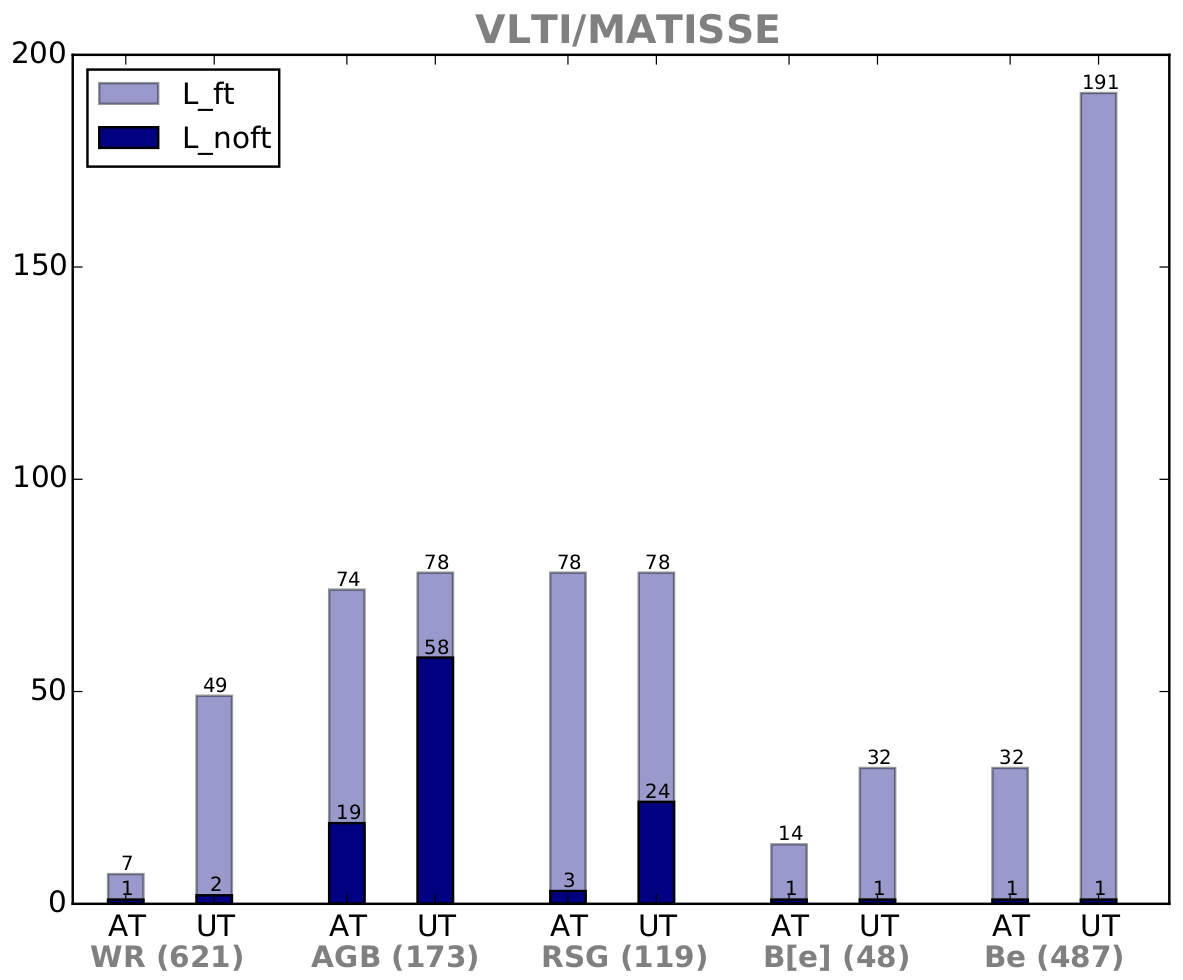}
    \end{tabular}
  \end{center}
  \caption[Number of targets observable] 
          { \label{fig:targets} {\bf Left:} Number of targets observable with MATISSE in low spectral resolution (R$\approx$35) for different types of stars, both in L band (\texttt{L\_noft}, dark blue bars) and N band (\texttt{N\_noft}, light blue bars). The transparent bars show the improvement of observability when using an external fringe tracker (\texttt{L\_ft} and \texttt{N\_ft}), like e.g. the foreseen project GRA4MAT. {\bf Right:} The same plot for the high resolution of MATISSE (R$\approx$4500 around the hydrogen Brackett $\alpha$ line).}
\end{figure}

The result of this study is shown in Fig.~\ref{fig:targets}. One can see that, out of our selected catalogs, roughly one third of the considered AGB targets, two third of the supergiant B[e] stars, and half of the considered red supergiant stars can be observed with MATISSE in its low spectral resolution mode (R$\approx$35). We are likely limited by our sample size in these cases.

On the other hand, very red targets like dusty Wolf-Rayet stars, or very blue targets like naked Wolf-Rayet stars or Be stars, will be more difficult to observe with MATISSE due to either the sensitivity limits of VLTI in the V band (very red targets), or to the MATISSE sensitivity limit (very blue targets). In both cases, the adjunction of a K-band fringe tracker to MATISSE, like the foreseen GRA4MAT project of ESO, will improve the situation by a large factor (especially when using the ATs).

That matter of fact is even more striking when using the high-spectral resolution, where we will be able to resolve the kinematic motion of gaz in the circumstellar envelopes\cite{}. From a handful of targets that will be observable with MATISSE alone, the adjunction of an external fringe tracker will allow the instrument to go at full throttle for the delivery of scientific results.

\section{Massive stars}

Massive stars are stars with a mass above 8 solar masses (\msun). They evolve fast and blast as supernovae at the end of their lives, not without having gone through a fireworks of different physical processes. Among them, one can count the main sequence O and B-type stars, red, yellow and blue supergiant stars, Wolf-Rayet stars, and other types of exotic stars. We invite the reader to take a look to the different other papers of this conference for more details\cite{Hofmann2016,Soulain2016}.

\subsection{Dusty blue supergiant stars}

Several blue supergiant stars (Wolf-Rayet, Luminous Blue Variable, supergiant B[e] stars) produce large amounts of dust before exploding as supernovae. Though they are not the main dust-producers in our Galaxy, they do vastly influence their surroundings by inputting kinetic energy and dust in their vicinity. In some cases, they can even trigger star formation.

The circumstellar environment (CE) around blue supergiant stars is often dense and highly anisotropic. This anisotropy could be driven either by fast-rotation and/or by the presence of a companion star. For example, fast-rotating B and O stars will present a denser and slower wind in the equatorial direction, and a lighter and faster wind at the poles, according to the usual sketch\cite{1991A&A...244L...5L}. At the equator, the density can be high-enough to veil the UV radiation from the star, allowing dust to condense into a disk-like structure, visible in the infrared. Such a break of symmetry of the CE has strong implications on the later phases of stellar evolution, as it can affect the stellar angular momentum and act as a brake for ejected material.

On the other hand, a secondary star orbiting around an O or a B star can affect the wind of the main star through gravitational influence, and concentrate material in the orbital plane. The consequence is similar as for rotating stars, i.e. an over-density in the equatorial plane that can trigger dust formation. However, the shape of such a disk-like structure would likely be different than in the first case, and there are opportunities of detecting the companion star, depending on the flux ratio between the two stars\cite{Millour2009a, Kraus2012}.

\begin{figure}[htbp]
  \begin{center}
    \begin{tabular}{cc}
      \includegraphics[width=1.\textwidth]{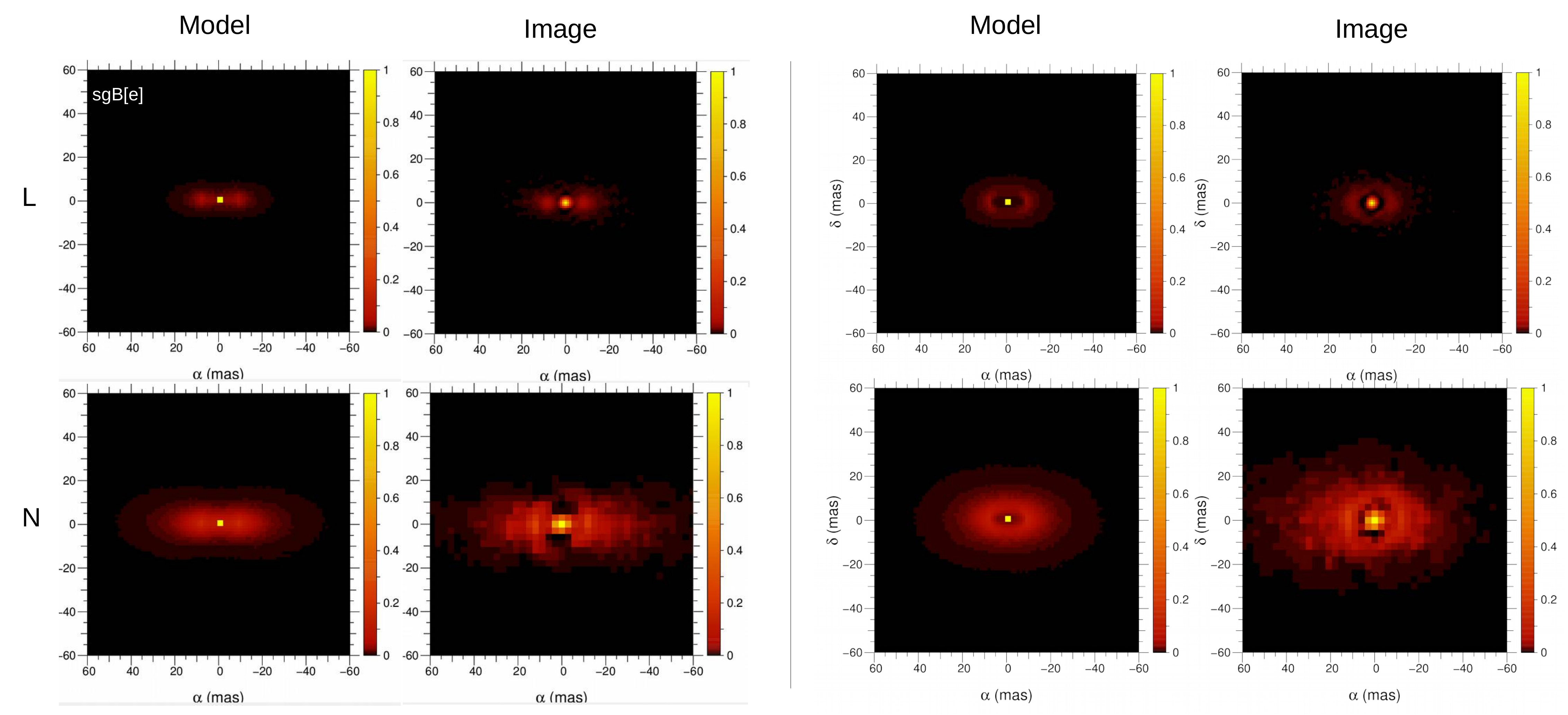}
    \end{tabular}
  \end{center}
  \caption[B\[e\] images] 
          { \label{fig:B_crochet_e} {\bf On the left}, one can see modelled ({\bf Left}) vs. reconstructed images ({\bf Right}) of typical B[e] star disks for a 90 degrees inclination (edge-on disk). {\bf The right panel} shows a 60 degrees inclination disk. The image reconstruction software used here is MiRA.}
\end{figure}

To prepare the MATISSE program on supergiant B[e] stars, we simulated what are the imaging performances of MATISSE in the case of a disk-like structure and a possible influence of fast-rotation of the hot star. To prepare and interpret the MATISSE observations we used a NLTE Monte Carlo radiation transfer code to compute a grid of supergiant B[e] models dedicated to mid- and near-IR interferometry \cite{Domiciano-de-Souza2012_v464p149}. The case of binarity is expected to be much easier, as it was already demonstrated that imaging with few baselines can be done in such a case\cite{Millour2009a, Kraus2012}. We simulated MATISSE with homemade scripts and produced OIFITS\cite{2005PASP..117.1255P} files. We input these files into the \texttt{MiRA} image reconstruction software developed by E. Thi\'ebaut\cite{Thiebaut2010a}. The results of our study is shown in Fig.~\ref{fig:B_crochet_e}. One can see that we are able to reconstruct most of the features seen in the model images: the central point source, the extent of the disk, and the inner hole seen in the 60 degrees inclination images. This study confirms that MATISSE will be a perfect imaging machine for these types of stars.

\subsection{The case of $\eta$ Carinae}

The Luminous Blue Variable (LBV) phase is a short-lived phase in the life of  massive stars.  The strong mass loss in this phase is not well understood. The LBV $\eta$ Car provides a unique laboratory for studying the  evolution of massive stars and the massive stellar wind during the LBV phase. $\eta$~Car is also an eccentric binary with a period of 5.54 yr.  X-ray studies demonstrate that there is a wind-wind collision zone, which is depends on the orbital phase.
Infrared interferometry with the VLTI allows us to study the stellar wind of the primary star and the wind-wind collision zone of the primary and secondary star with unprecedented  angular resolution of about 5\,mas and simultaneously with high spectral resolution. Using the VLTI, the diameter  of $\eta$~Car's wind region  was measured to be about 4.2\,mas in the $K$-band continuum\cite{2003A&A...410L..37V,2007A&A...464.1045K,2007A&A...464...87W}  (50\% encircled intensity diameter). In the He\,I\,2.059\,$\mu$m and the Br$\gamma$\,2.166\,$\mu$m emission lines, diameters of  6.5 and 9.6\,mas were measured, respectively\cite{2007A&A...464...87W}. 

For comparison, the radius of the primary star of $\eta$ Car is on the order of 100 solar radii $\sim$ 0.47\,au $\sim$ 0.20\,mas\cite{2001ApJ...553..837H}. This means that the measured 4.8\,mas Br$\gamma$ wind radius is about 25  times larger than the stellar radius. The measured line visibilities agree with predictions of the radiative transfer model of Hillier et al.\cite{2001ApJ...553..837H}. 
VLTI-AMBER observations  (performed at the beginning of 2014, about 5 to 7 months before the August 2014 periastron passage) allowed the reconstruction of velocity-resolved aperture-synthesis images  in more than 100 different spectral channels distributed across the Br$\gamma~$2.166\,$\mu$m  emission  line\cite{Hofmann2016, Weigelt2016}. The intensity distribution of the obtained images strongly depends on wavelength.  Interestingly, in the blue wing of the Br$\gamma~$2.166\,$\mu$m  emission  line, the wind region is much more extended than in the red wing.  At radial velocities of approximately $-$140 to $-$376\,km/s measured relative to line center (i.e., in the blue line wing), the intensity distribution is fan-shaped with a position angle of the  symmetry axis of  $\sim$126 degree. The fan-shaped structure extends approximately  8\,mas ($\sim$19\,au) to the south-east   and  6\,mas ($\sim$14\,au) to the north-west (measured at the 16\% intensity contour). 

Three-dimensional smoothed particle hydrodynamic simulations of $\eta$~Car's colliding winds  predict a very large wind density distribution that is more extended than the undisturbed primary star wind density distribution. At the time of our observations, the secondary star of $\eta$~Car was in front of the primary star, and therefore  the extended wind collision cavity was opened up into our line of sight. The comparison of the model images with the shape and wavelength dependence of the  intensity distributions  of the reconstructed VLTI images suggests that the obtained VLTI images are the first direct images of the innermost wind-wind collision zone.  

In the red line wing (positive radial velocities), the extension of the resolved wind structure is much smaller than in the blue wing, because the red-wing light is emitted from the wind region behind the primary star, where the primary wind is not disturbed by the wind collision (i.e., at positive radial velocities, we are looking through the optically thin wind collision zone in front of the primary). The $\eta$~Car channel maps provide  velocity-dependent image structures that can be used to test three-dimensional hydrodynamical  models of the massive interacting winds.

\subsection{Dust plumes around Wolf-Rayet stars}

We also investigated the potential of MATISSE to detect asymmetries in the dusty environment of massive blue stars, with an application to the detection of several spiral dust plumes around dusty Wolf-Rayet stars. These dust plumes likely trace the dust forming in the wind collision zone of a binary system composed of a WR and an O star. Many dust shells have been detected around Wolf-Rayet stars, but few of them were resolved into spiral plumes. MATISSE has the potential to image most of these dust shells at a few milliseconds of arc resolution, enabling the detection of many more spiral plumes, leading to a confirmation that WC8 and WC9 spectral types are linked to the inner system binarity.

\begin{figure}[htbp]
  \begin{center}
    \begin{tabular}{cc}
      \includegraphics[width=.6\textwidth]{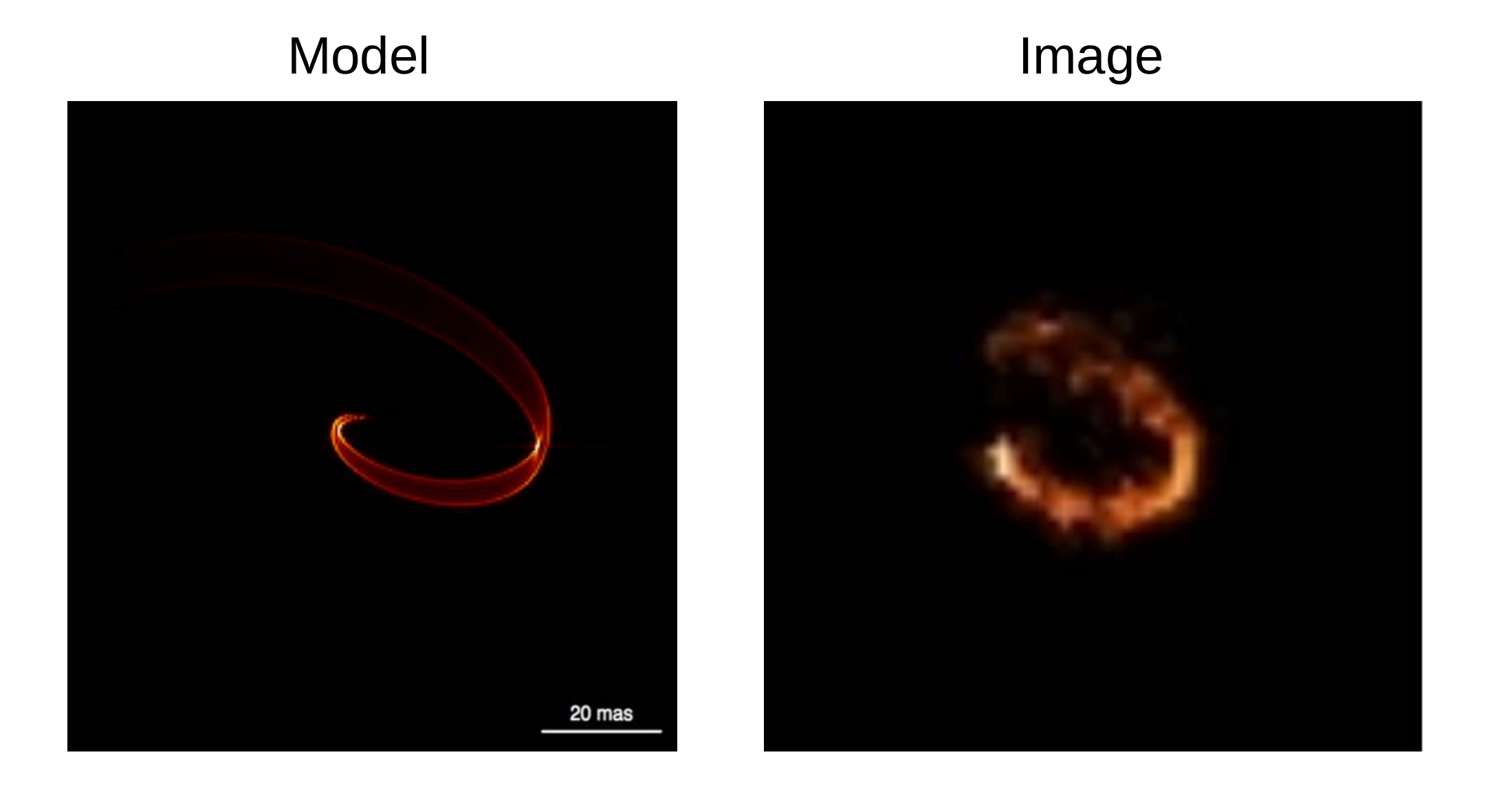}
    \end{tabular}
  \end{center}
  \caption[pinwheel image] 
          { \label{fig:pinwheel} Modelled (Left) vs. reconstructed images (Right) of a typical "pinwheel nebula" around a dusty Wolf-Rayet star. We used \texttt{IRBis} for this test case.}
\end{figure}

To illustrate the MATISSE potential, we simulated observations of a phenomenological "pinwheel" model with MATISSE using the readily available ASPRO software that has been updated to take into account the MATISSE-specific noises. The obtained OIFITS files were fed into the MATISSE official image reconstruction tool \texttt{IRBis}, which notably won the interferometry beauty contest of this year\cite{Sanchez2016}. The result is shown in Fig.~\ref{fig:pinwheel}, evidencing that MATISSE is indeed able to resolve a "pinwheel" nebula in 3 nights of observation.

\subsection{Young massive clusters}

Young massive clusters, are gravitationally bound groups of newly formed stars. These objects provide a rare opportunity to study the formation of massive stars and their evolution in the early stages. Among the well-known young clusters, R136 is located in the heart of 30Doradus (in LMC) and hosts so far the most massive stars known in the Local Universe. These stars are born in the large HII regions, embedded in gas and dust. MATISSE can be used to resolve the crowded core of R136-like clusters with higher angular resolution than conventional telescopes equipped with adaptive optics (e.g. SPHERE), and at longer wavelengths (L- and N-bands), in order to overcome the problem of interstellar extinction. 

\begin{figure}[htbp]
  \begin{center}
    \begin{tabular}{cc}
      \includegraphics[width=.6\textwidth]{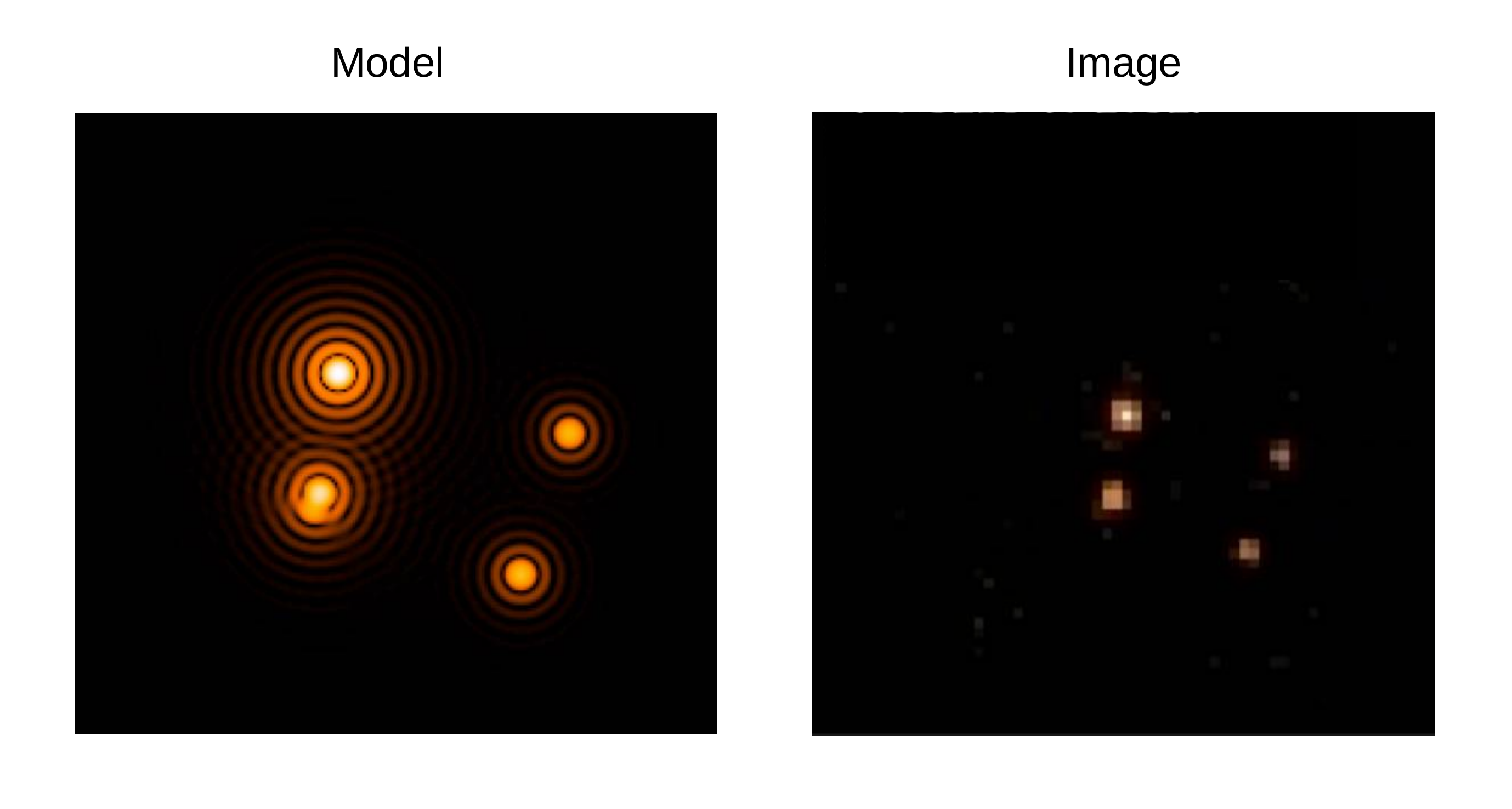}
    \end{tabular}
  \end{center}
  \caption[Cluster image] 
          { \label{fig:clusters} Simulated image of the core of a R136-like cluster at the age of 2 Myr in N-band ({\bf Left}) vs. reconstructed images ({\bf Right}). The left panel shows $0.3" \times 0.3"$ (FoV of MATISSE using UTs) covering $0.075\times0.075 pc^2$ of the core of the cluster at the distance of LMC (50 Kpc). The image reconstruction software used here is \texttt{IRBis}.}
\end{figure}

We simulated R136 using the publicly available N-body gravitational simulation \texttt{Nbody6} code\footnote{available at \url{http://www.ast.cam.ac.uk/~sverre/web/pages/nbody.htm}}. The left part of Figure \ref{fig:clusters} shows the inner $0.075 \times 0.075$\,pc of the simulated cluster, which correspond to the FoV of MATISSE using the UTs (0.3"). This snapshot of the core from the numerical \texttt{Nbody6} simulation is done at the age of 2 Myr in the N-band and at the distance of the LMC (50 Kpc).

We used the \texttt{ASPRO} tool from JMMC\footnote{available here: \url{http://www.jmmc.fr}} to simulate the MATISSE observables (visibilities, closure phases and differential phases), and the \texttt{IRBis} software to reconstruct the synthetic image. The result can be seen in the right panel of Figure \ref{fig:clusters}. The core massive stars have separations down to 24 mas. However, we also put a secondary companion in a binary system with separation of 3.5\,mas (seen close to the lower-left star in the left part of Figure \ref{fig:clusters}), and this companion star cannot be resolved by MATISSE in the N band. However, we did not perform image reconstruction tests in the L band, and it could be that this companion star is resolved at this wavelength, where MATISSE has a factor 3 higher angular resolution. Further tests need to be done to fully characterize the MATISSE performances in the context of star cluster observations.

\subsection{Red supergiant stars}

Red supergiants (RSGs) are strong contributors to the chemical and dust enrichment of the Galaxy through their mass-loss ($2\times 10^{−7} $M$_⊙ / $yr -- $3 \times 10^{−4} $M$_{⊙}$/yr, de Beck et al. 2010) and the coming blast as supernova. Mass-loss process still remains poorly understood. RSGs do not experience regular flares or pulsations and the triggering of their mass loss remains a mystery. It may be linked to the magnetic activity, the stellar convection, and the radiation pressure on molecules and/or grains, or all these together. The stellar surface and circumstellar dynamics and constitution of these stars have to be studied in detail to unveil the mass-loss mechanism.

MATISSE will open new insight at wavelengths still poorly observed for RSGs. The observations will range from the inner stellar photosphere (L and M band probing molecular transitions and pseudo- continuum) to the outer enveloppe (N band). The temporal variation is crucial  to make a step forward into the understanding of the stellar dynamics and the link with the mass-loss mechanism.

\section{Lower mass stars: AGB \& pulsating stars}

\subsection{AGB stars}

Most of the low and intermediate mass stars end their lives on the Asymptotic Giant Branch (AGB) phase. During this phase, pulsation and radiation pressure on dust leads to a phase of strong mass loss, during which gas and dust enriched by the products of the star's nucleosynthesis will be ejected. This mass loss is crucial for the chemical enrichment of the ISM and therefore for the chemical evolution of galaxies. A pivotal aspect of the mass-loss process is its geometry, i.e. the density distribution of the circumstellar envelope of the AGB stars at different scales and different evolutionary phases.

To understand the mass-loss process, it is essential to study the mass-loss from very deep inside the star up to the interface with the ISM. Due to its broad wavelength coverage MATISSE is the unique instrument to study the different dust and molecular species present in the atmospheres and envelopes of AGB stars. The aim is to achieve a complete view from the upper photosphere to the outer envelope layers and to complement Herschel and MIDI surveys (see Paladini et al., this conference).

\subsection{Envelopes around Cepheid stars}

A powerful way of constraining the period-luminosity (PL) relation is to use the Baade-Wesselink (BW)
method of distance determination. The basic principle of the BW method is to compare the
linear and angular size variation of a pulsating star in order to derive its distance through
a simple division. The angular diameter is either derived by interferometry\cite{kervella04a} (hereafter IBW
for Interferometric Baade-Wesselink method), or by
using the InfraRed Surface Brightness (IRSB) relation\cite{storm11a,storm11b}. However,
when determining the linear radius variation of the Cepheid by spectroscopy, one has to use
a conversion projection factor from radial to pulsation velocity \cite{nardetto04,merand05}. In this method, angular and linear diameters have to correspond to the same physical layer in the star to provide a correct estimate of the distance\cite{nardetto07,nardetto09}. Consequently, the circumstellar environment (CSE) around Cepheids should be considered in both versions of the BW method, particularly when deriving the angular diameter curve. The impact of CSE on the period-luminosity relation of Cepheids need also to be established. 

\begin{figure}[htbp]
  \begin{center}
    \begin{tabular}{cc}
      \includegraphics[width=.6\textwidth]{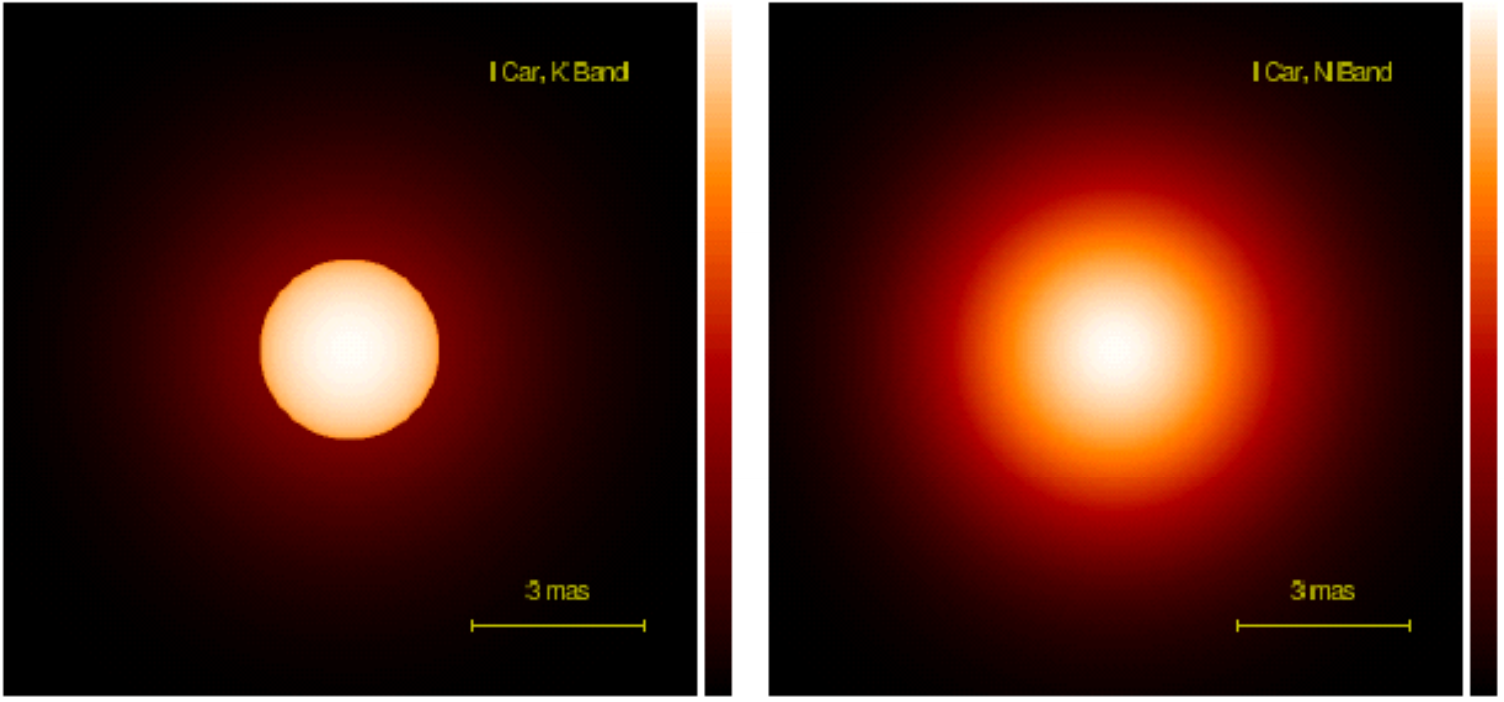}
    \end{tabular}
  \end{center}
  \caption[lcar] 
          {\label{lcar}{\bf Left:} the geometry of the surrounding envelop of l Car (in K band) as constrained by VINCI observations (Kervella et al. 2006). The envelope is supposed centro-symmetric and the Full-Width at Half-Maximum of the modelled gaussian is around 3 stellar radii and a flux contribution of 4\%. {\bf Right:} the geometry of the envelope expected in the N band is almost the same but with a larger flux contribution (figure taken from Antoine Merand PhD).}
\end{figure}

Envelopes around Cepheids have been discovered by long-baseline interferometry in the K Band with VLTI\cite{kervella06a} and CHARA\cite{merand06} (see Fig.\ref{lcar}). Since then, four Cepheids have been observed in the N band with VISIR and MIDI\cite{kervella09,gallenne13b} and one with NACO\cite{gallenne11,gallenne12}. Some evidences have also been found using high-resolution spectroscopy\cite{nardetto08b}. The size of the envelope seems to be at least 3 stellar radii and the flux contribution in K band is from 2\% to 10\% of the continuum, for medium- and long-period Cepheids respectively, while it is around 10\% or more in the N band (estimated also from SED derived with VISIR). 

Recently, Nardetto et al. (2016, in press) found a resolved structure around the prototype classical Cepheid delta Cep in the visible spectral range using visible VEGA/CHARA interferometric data observations. These data are indeed consistent in first approximation with a quasi-hydrostatic model of pulsation surrounded by a static CSE with a size of  8.9 $\pm$ 3.0 mas and a relative flux contribution of 0.07 $\pm$ 0.01. A model of visible nebula (a background source filling the field of view of the interferometer) with the same relative flux contribution is also consistent with the data at small spatial frequencies. However, in both cases, they find discrepancies in the squared visibilities at high spatial frequencies (maximum 2$\sigma$) with two different regimes over the pulsation cycle of the star, $\phi=0.0-0.8$ and $\phi=0.8-1.0$. One possibility could be that the star is lighting up its environment differently at minimum ($\phi=0.0-0.8$) and maximum ($\phi=0.8-1.0$) radius. This reverberation effect would then be more important in the visible than in the infrared since the contribution in flux of the CSE (or the background) in the visible band is about 7\% compared to 1.5\% in the infrared (Merand et al. 2015).

The processes at work in infrared and in the visible regarding the CSE are different. We expect thermal emission in the infrared and scattering in the visible. MATISSE offers an unique opportunity to study the envelopes of Cepheids. Determining the size of these envelope (as a function of the spectral band) and their geometry (as a function of the pulsation phase), will bring insights in the links between pulsation, mass loss and envelopes.

\acknowledgments     

This research has made use of the SIMBAD database,
operated at CDS, Strasbourg, France. We used data from the Infrared Space Observatory (ISO) from ESA, and from the Infrared Astronomical Satellite (IRAS).

We are grateful to ESO, CNRS/INSU, and the Max-Planck Society for continuous support in the MATISSE project. 

NN acknowledges the support of the French Agence Nationale de la Recherche (ANR), under grant ANR-15-CE31-0012-01 (project UnlockCepheids). NN acknowledges financial support from “Programme National de Physique Stellaire” (PNPS) of CNRS/INSU, France.


\bibliography{main}   
\bibliographystyle{spiebib}   

\end{document}